
\input jnl
\oneandfourfifthsspace
\def\sperp{{\scriptscriptstyle\perp}}
\preprintno{FTG-110-USU}\dateline
\title Is General Relativity an ``Already Parametrized'' Theory?
\author C. G. Torre
\affil Department of Physics
Utah State University
Logan, UT  84322-4415
USA
\abstract
\endtitlepage
Beginning in the late fifties and early sixties with the work of Dirac
\refto{1} and
Arnowitt, Deser, and Misner \refto{2} the canonical dynamical structure
of
general relativity has often been viewed as that of a ``parametrized field
theory'': a generally covariant version of any field theory
obtained by adjoining many-fingered spacetime variables---spacelike
embeddings and their conjugate momenta--- to the original set of canonical
field variables.  A decade later Kucha\v r extensively developed the
structure of
parametrized field theories and their role as a paradigm for
canonical gravity \refto{3, 5}.   Essentially, the idea behind this work is
that
the
embedding variables are
somehow camouflaged amongst the canonical coordinates and momenta of
geometrodynamics,  and as these variables define (internally specified)
spacetime points, their canonical conjugates must necessarily be
identified with energy and momentum densities for the true degrees of
freedom of the gravitational field.  From the point of view of parametrized
field theory this identification constitutes the
meaning of the Hamiltonian and momentum constraints which appear as a
subset of the Einstein equations.

It is not hard to see why this paradigm
was so attractive to so many workers in general relativity.  Classically, if
canonical gravitational dynamics can be cast into the mold of parametrized
field theory then one has a clean separation between kinematical (or ``pure
gauge'') aspects of general relativity and the truly dynamical aspects. In
particular the Cauchy problem becomes especially simple because the
constraints reduce to conditions on the embedding momenta and it is then
quite clear what are the freely specifiable initial data.

While the parametrized field theory paradigm leads to a pleasantly simple
picture of classical dynamics of the gravitational field, it is when trying to
understand the quantum mechanics of the gravitational field that this
paradigm becomes extremely attractive.  If the embeddings and their
conjugate momenta can be gleaned from the phase space of general
relativity then it is straightforward, at least formally, to construct a
resolution of the well-known problems of time and observables.  Briefly
stated, the problem of time is concerned with the absence of any structure
that can be taken as time for the purposes of quantization. If
the many-fingered spacetime variables can be isolated then one has
internally
specified ``rods and clocks'' that can be used for the construction and
interpretation of quantum theory.  Closely related to the problem of time is
the problem of observables which arises from the fact
that not a single observable, \ie function on the constraint surface that is
invariant under the canonical transformations generated by the constraints,
is known in general relativity (in the case of a closed universe).   This
spells difficulty for interpretation of quantum gravity formulated {\it a la}
Dirac, where one is to select physical states of the gravitational field by
demanding they satisfy operator versions of the Hamiltonian and
momentum constraints.  In this approach to quantization it is only the
observables that have a chance to be promoted to (self-adjoint)
operators on the
physically allowed (Hilbert) space of states.  In the quantum version of the
parametrized formalism, the constraints
become functional Schrodinger equations and it is then
consistent to take the observables to be (functions of) the freely specifiable
Cauchy data.  Wavefunctions satisfying the functional
Schrodinger equations represent probability amplitudes for measuring the
free Cauchy
data on a hypersurface set by the embedding variables.

Despite decades of effort, it has proven a very difficult task to implement
the point of view that general relativity is a parametrized field theory.  The
best attempt in the context of the vacuum theory is related to the conformal
approach to the initial value problem \refto{4} in which the role of
many-fingered time is given to the mean extrinsic curvature of the
hypersurface
upon which the canonical data are defined.  However, despite the great
utility of this formalism for attacking the Cauchy problem in general
relativity, it is not yet adequate for solving the fundamental problems of
canonical quantum gravity \refto{5}.  If
one allows coupling to matter, the situation can improve rather
dramatically
\refto{6}, but even when using matter to give invariant meaning to
spacetime
points there are still some drawbacks \refto{5}.

A good deal of effort has been devoted to the study of simple models that
allow for the implementation of the parametrized theory paradigm with
varying degrees of success \refto{5}, and  such
investigations expose the difficulties that can occur.  But, because one
is dealing with models, one is never sure exactly what is going to be the
situation in the full theory.  The current state of affairs is that very
little is known about the possibility of transforming the full vacuum theory
into the parametrized formalism.

The purpose of this essay is effectively to place
an upper bound on the degree to which one can view general relativity as a
parametrized field theory, and discuss the implications for canonical
quantization based on this paradigm.  More precisely, we will point out that
there is an obstruction to finding a bijection that would identify
the constraint surface of general relativity with that of any parametrized
field
theory.  The proof relies heavily on results and techniques of Arms,
Fischer, Isenberg, Marsden and Moncrief \refto{7} and is essentially a {\it
reductio ad absurdum}.  Throughout we confine attention to closed
universes, \ie the spacetime manifold is $M=R\times\Sigma$ with $\Sigma$
compact.

Let  $(\Gamma, \Omega)$ denote the phase space and symplectic structure
for general relativity. $\Gamma$ is taken
to be a (suitably defined \refto{7}) cotangent bundle over the space of
3-metrics on a compact three-dimensional manifold $\Sigma$ and
$\Omega$ is its canonical symplectic structure.
Dynamically admissible points in $\Gamma$ lie on the constraint
surface $\bar\Gamma\subset\Gamma$, which satisfies the Hamiltonian and
momentum constraints:
$$H_\sperp=0=H_a .$$
Let $(\Upsilon, \omega)$
denote the phase space of a parametrized field theory and its symplectic
structure. $\Upsilon$ is the product
of the cotangent bundle over embeddings of $\Sigma$ into the
spacetime manifold $M$ (equipped with the canonical symplectic structure)
and the phase space for a field theory, which is taken to be an
infinite-dimensional
symplectic manifold. Denote points in the phase space for the field theory
as  $Z^A$, which are a collection of fields on
$\Sigma$.  Let $Q:\Sigma\to M$ represent an embedding, and denote the
momentum conjugate to the embedding
as $P_\alpha$, which is geometrically a cotangent vector to the space of
embeddings.  Dynamically allowed points in $\Upsilon$ also lie on a
constraint surface $\bar\Upsilon\subset\Upsilon$:
$$\Pi_\alpha:=P_\alpha+h_\alpha(Q,Z^A)=0.$$
Here $h_\alpha$ has the physical interpretation as the flux of
energy-momentum of the fields $Z^A$ through the hypersurface defined
by
$Q$.

We can now make precise the sense in which one would like to view
general relativity as a parametrized field theory.

\noindent {\it Conjecture:}  There is a bijection
$\Phi:\Upsilon\to\Gamma$ which identifies the constraint surface of
general relativity with that of a parametrized field theory:
$$\Phi(\bar\Upsilon)=\bar\Gamma.\tag$$

The conjecture amounts to the statement that there is a change of variables
on the phase space of general relativity such that, using the Hamiltonian and
momentum constraints,  $4\infty^3$ of the new variables can be solved for
in terms of the remaining variables.
In practice one would like to make the conjecture a good bit stronger.
First, one would like to demand that the map $\Phi$ is a {\it
diffeomorphism}; further, this diffeomorphism should identify
the symplectic structures on $\Gamma$ and $\Upsilon$, \ie
$\omega=\Phi^*\Omega$.  This means that the change of variables
mentioned above is in fact a canonical transformation (this requirement can
be weakened somewhat).   Second,  it is necessary that the embeddings
$Q$ are spacelike with respect to the Einstein metrics which solve the
Hamiltonian equations of motion.    These {\it addenda} to the above
conjecture
are irrelevant for our present purposes because the
conjecture already fails without the added assumptions.

The conjecture fails because $\bar\Gamma$ is a stratified manifold,
\ie  it has ``conical singularities'' at phase space points representing
Cauchy data for spacetimes with Killing vectors \refto{7}.  It is precisely
these
singular points which prevent one from solving the constraints as indicated
by the conjecture.  More precisely,  it can be shown that $\bar\Upsilon$ is
everywhere a smooth manifold, and hence there cannot be a
bijection that identifies $\bar\Upsilon$ and $\bar\Gamma$.  The
conjecture is false.

The proof that $\bar\Upsilon$ is globally a manifold follows a model
calculation of Arms \refto{8}.  Let ${\cal E}$ represent the space of
cotangent vectors to the space of embeddings.  Define the map
$\Pi:\Upsilon\to {\cal E}$ via $\bar\Upsilon=\Pi^{-1}(0)$.  If the
differential $d\Pi$, which is a linear map from the tangent space at a point
of $\Upsilon$ to ${\cal E}$, is surjective at
a point in $\bar\Upsilon$, then there is a neighborhood
$U\subset\bar\Upsilon$  of that point which is a smooth submanifold of
$\Upsilon$.
Surjectivity follows if it can be shown that the natural adjoint
of $d\Pi$ is injective and has injective symbol everywhere on
$\bar\Upsilon$ \refto{7}.  It is straightforward to verify that at each point
of
$\Upsilon$ the symbol of the adjoint operator is injective (in the
generalized sense of \refto{9})
as is the adjoint operator itself, and hence the map
$d\Pi$ is surjective everywhere on $\Upsilon$. This
guarantees that $\bar\Upsilon$ is everywhere a manifold.

So it seems that the best one can hope for in general relativity is to find an
identification of gravitation as a parametrized field theory only for the
generic spacetimes, \ie those without symmetry.  To put this result
into a proper perspective we must see (i) what assumptions in the proof
can be sensibly relaxed, and (ii) what it implies in practice, \eg for
canonical quantum gravity.   While such an investigation is obviously
beyond the possible scope of this essay, we can make a few comments here.

It is tempting to think that one should relax the assumption that the phase
space of fields $Z^A$ is a (symplectic) manifold because this would allow
for the possibility that both $\bar\Upsilon$ and $\bar\Gamma$ are
manifolds with singularities.  The problem with this strategy is that,
because $\Gamma$ is a manifold, it is now impossible to find a
diffeomorphism that identifies $\Gamma$ and $\Upsilon$, and so this
obvious modification of the conjecture is still problematic.

A more intriguing way to get around the obstruction encountered above is
to redefine the gravitational phase space; a prime example of this being
Ashtekar's  Hamiltonian formulation of general relativity \refto{10}.
Here one can
attempt to identify the constraint surface of general relativity with that of a
parametrized {\it gauge theory}.  Because the constraint surface for a
gauge
theory (even without parametrization) is not
necessarily a manifold \refto{11}, the argument given above does not need
to
apply.
The formulation of parametrized gauge theory in general and particularly
its use as a paradigm for Ashtekar's formulation of general relativity are
topics worthy of further investigation.

In the spirit of trying to find a phase space formulation of general
relativity that is more amenable to the parametrized field theory
paradigm, it is worth pointing out that the ``covariant phase space''
approach to gravitation \refto{12} is available.  In this picture the putative
relation
with parametrized field theory may change considerably.  Indeed, the
covariant phase space formulation of parametrized field theory is
substantially
different from its canonical counterpart under consideration here
\refto{13}.  This
is another topic worthy of further investigation.

Let us conclude by discussing the quantum mechanical implications of the
obstruction to the above conjecture.   We know that there is an
obstruction to identifying $\bar\Gamma$ with $\bar\Upsilon$, but we do
not know what actually fails when trying to make the
identification except in simple models.  Two (very similar) models can be
exhibited which nicely illustrate the type of obstruction we are discussing.
One is a model of Kucha\v r \refto{16, 8, 17, 15} the techniques of which
are easily
generalized \refto{14, 18} to the ``midi-superspace'' of spacetimes that
admit
two
commuting Killing vectors (compact spatial topology in both cases).  The
key feature of these models is that it is possible to find a map
that satisfies (1), but this map is not bijective.  Alternatively, one
can find a bijection between the gravitational model\footnote*{Actually, in
this case one embeds the model gravitational phase space in a (slightly)
larger phase
space, much as we suggested might be useful using, \eg the Ashtekar
formalism.  } and the model
parametrized theory, but the gravitational constraint surface appears as a
submanifold (with singularities) of the constraint surface of the
parametrized field theory (so that (1) is violated).  As shown in \refto{15},
in the
quantum theory this
latter possibility leads to a set of functional Schrodinger equations, and all
the advantages associated with such equations, along with a finite number
of
subsidiary conditions to be placed on the quantum mechanical state vector.
Therefore, in these models, the physical interpretation allowed by the
parametrized field theoretic structure remains intact, but the notion of
observable has to be modified slightly to be consistent with the subsidiary
conditions.
If this situation
could be shown to persist in the full theory, the failure of the conjecture
would not necessarily preclude the use of the paradigm to resolve the
difficult conceptual and technical issues facing canonical quantum gravity.

I would like to thank Abhay Ashtekar and Karel Kucha\v r for discussions.

\references
\refis1{P.A.M. Dirac, {\it Lectures on Quantum Mechanics}, (Yeshiva
University, New York, 1964).}

\refis2{R. Arnowitt, S. Deser, and C. Misner in {\it Gravitation: An
Introduction to Current Research}, edited by L. Witten (Wiley, New York
1962).}

\refis3{K. V. Kucha\v r, \jmp 13, 758, 1972; \jmp 17, 801, 1976.}

\refis4{See, \eg Y. Choquet-Bruhat and J. York in {\it General Relativity
and Gravitation: 100 Years After the Birth of Albert Einstein, Vol. 1},
edited by A. Held (Plenum, NY 1980).}

\refis5{K. V. Kucha\v r, ``Time and Interpretations of Quantum Gravity'',
to
appear in the proceedings of {\it The Fourth Canadian Conference on
General Relativity and Relativistic Astrophysics}, edited by G. Kunstatter,
D. Vincent, and J. Williams (World Scientific, Singapore 1992).}

\refis6{K. V. Kucha\v r and C. G. Torre, \prd 43, 419, 1990; \prd 44,
3116, 1991.}

\refis7{See A. Fischer and J. Marsden in {\it General Relativity: An
Einstein Centenary Survey}, edited by S. Hawking and W. Israel
(Cambridge University Press, Cambridge 1979); J. Isenberg and J.
Marsden, \prpts 89, 181, 1982, and references therein.}

\refis8{J. Arms, \jmp 21, 15, 1980.}

\refis9{L. H\" ormander, \journal Ann. Math., 83, 129, 1966.}

\refis{10}{See A. Ashtekar, {\it Lectures on Non-Perturbative Canonical
Gravity}, (World Scientific, Singapore 1991) and references therein.}

\refis{11}{J. Arms, \jmp 20, 443, 1979.}

\refis{12}{C. Crnkovic and E. Witten in {\it 300 Years of Gravitation},
edited by S. Hawking and W. Israel (Cambridge University Press,
Cambridge 1987);
J. Lee and R. Wald, \jmp 31, 725, 1990; A. Ashtekar, L. Bombelli, and O.
Reula, in {\it Mechanics,
Analysis and Geometry : 200 Years After Lagrange}, edited by M.
Francaviglia ( North-Holland, New York 1991). }

\refis{13}{C. G. Torre, ``Covariant Phase Space Formulation of
Parametrized Field Theories'', Utah State University Preprint, 1992.}

\refis{14}{C. G. Torre, in preparation, 1992.}

\refis{15}{K. V. Kucha\v r and C. G. Torre in {\it Conceptual Problems
of Quantum Gravity}, edited by A. Ashtekar and J. Stachel, (Birkh\"auser,
Boston 1991).}

\refis{16}{K. V. Kucha\v r, \jmp 19, 390, 1978.}

\refis{17}{K. V. Kucha\v r, C. G. Torre, \jmp 30, 1769, 1989.}

\refis{18}{For the case in which the Killing vectors are hypersurface
orthogonal and the universe is open see K. V. Kucha\v r, \prd 4, 955,
1971;  C. G. Torre, \cqg 8, 1895, 1991.}

\endreferences
\endit